\documentclass[twocolumn,superscriptaddress,showpacs,aps,prb]{revtex4}
\usepackage{makeidx}
\usepackage{bm}
\usepackage{graphics}
\usepackage[dvips]{epsfig}
\newcounter{fig}

\begin{document}
\title{Cyclotron resonance  and Faraday rotation in graphite }
\author{L.A. Falkovsky}
\affiliation{L.D. Landau Institute for Theoretical Physics, Moscow
119334, Russia} \affiliation{Institute of the High Pressure
Physics, Troitsk 142190, Russia}
\pacs{71.20.Di, 76.40.+b, 78.30.-j}

\date{\today}      

\begin{abstract}
 The optical conductivity of graphite in quantizing magnetic fields is analytically evaluated for frequencies in the
range of 10--300 meV, where the electron relaxation processes can
be neglected and the low-energy excitations at the "Dirac lines"
are more essential.  The conductivity peaks are explained in terms of
 the electron transitions in graphite. 
 Conductivity calculated per one
graphite layer tends  on average to  the universal conductivity of graphene.
 The (semi)metal-insulator transformation is possible under doping in high  magnetic fields.
\end{abstract}
\maketitle
Graphite is usually  considered as a layered semimetal  composed of the graphene monolayers. Within this assumption,  the graphite electron spectrum was evaluated many years ago within the Slonczewski--Weiss--McClure (SWM) theory \cite{SW}. The  so called "Dirac cone" turns   into   four  bands with the twofold degenerate zero mode  where
electrons and holes are located \cite{Sem}. The zero-mode dispersion   is very small, of the order of 20 meV, in the main-axis direction  because  the interlayer interaction is weak. Unusual properties of graphite have attracted much attention for more than 50 years. The most accurate method to study the band structure of graphite is a study of the Landau levels (LLs)  through experiments such as magneto-optics \cite{TDD,DDS,LTP,OFM,OFS,OP,CLW} and magnetotransport  \cite{KTS,LK,JZS,SOP,RM}. However, the interpretation of the experimental results involves a significant degree of uncertainty since, as noticed by Doezema {\it et al}
\cite{DDS}
and Chuang {\it at al} \cite{CBN} "it is not clear where the resonances are to be marked." 

The SWM theory requires
the use of many tight-binding parameters and provides the simple description of observed phenomena either in the semiclassical limit of week magnetic fields or for high frequencies when the largest tight-binding parameter $\gamma_1 = 0.4$ eV plays the  leading role \cite{Fal}. At the relatively strong magnetic fields $B\sim$ 1--30 T and frequencies $\omega\sim$ 10--350 meV, the smaller tight-binding parameters $\gamma_2,
\gamma_5,$ and $\Delta$  of the order of 20 meV are essential. In this case, any physical property for graphite in magnetic field is represented by an integral over the momentum projection $k_z$. The SWM model can be simplified assuming that only the integration limits  produce the main contributions \cite{OFS,CBN,LA}. Such approximation is similar to the theory of magneto-optical effects in topological insulators \cite{TM} and graphene \cite{MHA}.  However, in the 3d systems, the other features 
such as the band extrema or the integration limits at the Fermi level   can contribute as well. Therefore,
the analytical expression for the dynamic conductivity in
the presence of magnetic fields is needed for an interpretation of  magneto-optics experiments. The theoretical study of magneto-optical properties in multilayer graphene
is realized in Ref. \cite{KA}.

In this paper, we evaluate a formula for the optical conductivity of graphite in the presence of  quantizing magnetic fields and results are compared with experiments. We remind the notation for the LLs in graphite 
using the Hamiltonian in the form of Refs. \cite{PP,GAW}.
The expression for both the longitudinal and Hall dynamical conductivities is given. The (semi)metal--insulator transition is discussed in  conclusions.

Neglecting the trigonal warping $\gamma_3$, the effective Hamiltonian near the $ KH-$line of the Brillouin zone can be written in the form
\begin{equation}
H(\mathbf{k})=\left(
\begin{array}{cccc}
\tilde{\gamma}_5 \,    & vk_{+} \,& \tilde{\gamma}_1    \, & 0\\
vk_{-} \,& \tilde{\gamma}_2     \, & 0\,& 0\\
\tilde{\gamma}_1    \,  &0 \,& \tilde{\gamma}_5 \,  &vk_{-}\\
0 \,& 0 \,&vk_{+} \,&\tilde{\gamma}_2
\end{array}%
\right)   \label{ham}
\end{equation}%
where $k_{\pm}=\mp ik_x-k_y$, $v=1.02\times10^8$ cm/s is the intra-layer velocity, and 
$\tilde{\gamma}_j$ are the functions of $k_z$:
\begin{eqnarray}\tilde{\gamma_1}=2\gamma_1\cos{(k_zc_0)}\,,
\tilde{\gamma}_2=2\gamma_2\cos{(2k_zc_0)}\,,\nonumber\\ \nonumber  \tilde{\gamma}_5=2\gamma_5\cos{(2k_zc_0)}
+\Delta
\end{eqnarray}
with the distance $c_0=3.35$ \AA\, between  layers in graphite.

At the magnetic field $B$, the momentum projections $k_{x,y}$ become the operators with the commutation rule $\{k_{+},k_{-}\}=-2e\hbar B /c$, and we can use the relations
\[k_+=\sqrt{2|e|\hbar B/c}\,a, \quad  k_-=\sqrt{2|e|\hbar B/c}\,a^+\,,\]
involving the creation and annihilation operators.
We seek the eigenfunction of the Hamiltonian  in the form
\begin{equation}
\psi_{sn}^{\alpha}(x)=
\left\{\begin{array}{c}
C^{1}_{sn}\varphi_{n-1}(x)\\
C^{2}_{sn}\varphi_{n}(x)\\
C^{3}_{sn}\varphi_{n-1}(x)\\
C^{4}_{sn}\varphi_{n-2}(x)\,
\end{array}\right.\label{func}
\end{equation}
where  the Landau number $n\ge2$ and $\varphi_{n}(x)$ are 
orthonormal Hermitian   functions with $n\ge0$. Then, every row
in the Hamiltonian (\ref{ham}) becomes proportional to the Hermitian function with the definite $n$, and we obtain 
a system of the linear equations for the eigenvector ${\bf C}_{sn}$
\begin{equation}
\left(
\begin{array}{cccc}
\tilde{\gamma}_5-\varepsilon \,    & \omega_c\sqrt{n} \,& \tilde{\gamma}_1    \, & 0\\
\omega_c\sqrt{n} \,& \tilde{\gamma}_2-\varepsilon     \, & 0\,& 0\\
\tilde{\gamma}_1    \,  &0 \,& \tilde{\gamma}_5-\varepsilon \,  &\omega_c\sqrt{n-1}\\
0 \,& 0 \,&\omega_c\sqrt{n-1} \,&\tilde{\gamma}_2-\varepsilon
\end{array}%
\right) \times\left(\begin{array}{c}
C^{1}_{sn}\\
C^{2}_{sn}\\
C^{3}_{sn}\\
C^{4}_{sn}
\end{array} \right)=0\,  \label{ham1}
\end{equation}
 where $\omega_c=v\sqrt{2|e|\hbar B/c}$.  
 \begin{figure}[]
\resizebox{.5\textwidth}{!}{\includegraphics{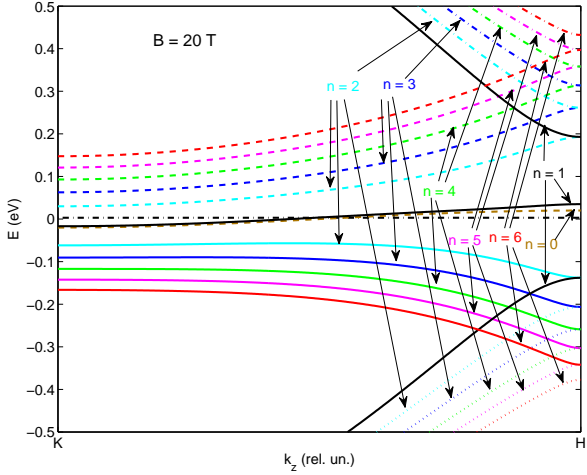}}
\caption{(Color online) LLs $\varepsilon_{sn}$ for $n$ = 0 to 6 in four bands $s$ as functions of momentum $k_z$ along the $KH-$line in the Brillouin zone 
($K=0,\, H = \pi/2c_0$) at  the magnetic field $B$ = 20~T using the SWM model with the TB parameters taken from Refs. \cite{PP, GAW} and neglecting trigonal warping;  the Fermi energy $\varepsilon_F=0$ eV shown in dash-dotted line. Electron transitions are possible between the levels  $\varepsilon_{s n}<0$   and $\varepsilon_{s'n'}>0$ with the selection rule  $\Delta n =\pm 1$\,. }
\label{disp20}
\end{figure}


  At each Landau number $n$, the eigenvalues
 $\varepsilon_{sn}$  of the Hamiltonian are marked by  the index of the band $s=1,2,3,4$   numerating  the levels from the bottom. We will use the notation $|sn\rangle$ for levels.  LLs with $n\ge 2$ are in every four bands,  shown in Fig.
 \ref{disp20} as functions of $k_z$ along the $KH-$line of  the Brillouin zone.  
 The eigenvalues are determined by the equation
 \begin{eqnarray}\nonumber
[(\tilde{\gamma}_5-\varepsilon)(\tilde{\gamma}_2-\varepsilon)-\omega_c^2(n-1)]\\ \nonumber
\times[(\tilde{\gamma}_5-\varepsilon)(\tilde{\gamma}_2-\varepsilon)-\omega_c^2 n]- \tilde{\gamma}_1^2(\tilde{\gamma}_2-\varepsilon)^2=0\,.\label{ll}\end{eqnarray}
    
 In addition, there are four  levels. One of them, $\varepsilon_{0}=\tilde{\gamma}_2$, with $n=0$   and the eigenvector ${\bf C}_0=(0,1,0,0).$ 
  This level  intersects   the Fermi level
 and belongs to the  electron (hole) band near the $K$     $(H)$ point.  Other  three levels, $s=1,2,3,$ are indicated by $n=1$  with
 $C^4_{s1}=0$. One level, $|21\rangle$, is
 very close to the level with $n=0$. This pattern is consistent with Ref. \cite{OFM}.
 The level structure at the $K'H'-$line is similar, therefore each level is fourfold degenerate, twice in spin and twice in pseudo-spin of the $KH$ and $K'H'$ valleys. 
 
 In the region, where $\tilde{\gamma}_1\gg \tilde{\gamma}_2,\tilde{\gamma}_5$, the two closest bands are written as
\[\varepsilon_{2,3}(n)=\tilde{\gamma_2}\pm \omega_c^2\sqrt{n(n-1)}/\tilde{\gamma}_1\,.\]

The Fermi energy  at zero temperature, $T$=0, is determined  by equality of electron and hole concentrations given by the sum of integrals 
\begin{equation}
n_{h,e}=\frac{\hbar\omega_c^2}{\pi^2v^2c_0}\sum_{sn}\int dz
\label{en}\end{equation}
over the region  in the Brillouin half-zone $ z=c_0 k_z$ where electrons and holes are located. As known, the Fermi energy oscillates in the quantum Hall regime. We will consider the relatively strong magnetic fields where
the oscillations do not exceed of 2 meV according to the  
 electro-neutrality condition (\ref{en}).
 
 At finite temperatures the conductivity is expressed in terms of the correlation function \cite{AGD} (for details see Ref. \cite{FV})
\begin{equation}
\mathcal{P}\left( \omega
\right) =T\sum\limits_{\omega
_{m}}\int dx dx'Tr\left\{ v^{i}%
\mathcal{G}\left( \omega_{+},x,x'\right) v^{j}\mathcal{G}\left( \omega_{-},x',x\right)
\right\}\,, \label{cor} \end{equation}
where $\mathcal{G}\left( \omega_{+},x,x'\right)$ is the temperature
Green's function, $\omega_{\pm}=\omega\pm\omega_m$\,,
$Tr$ is taken over the $|sn\rangle -$eigenstates, and the $x,x'-$integration is over the coordinate involved explicitly in the Hamiltonian while the Landau gauge is used. Then, the 
Fourier transform  of  the eigenstates with respect the $y-$coordinate is assumed.
 Using the eigenfunctions (\ref{func}), we write the
 Green's function of the Hamiltonian (\ref{ham}) 
 \[ \mathcal{G}^{\alpha \beta}(\omega,x,x')=\sum_{sn}\frac{\psi^{\alpha}_{sn}(x) \psi^{*\beta }_{sn}(x')}{i\omega-\varepsilon_{sn}}\,.\]
 
The intralayer-velocity matrix $v^{i}$ is given by the derivative of the  Hamiltonian (\ref{ham})
\begin{equation}
\mathbf{v}=\frac{\partial H(\mathbf{k})}{\partial \mathbf{k}}\,.
\label{vel}
\end{equation}%

The straightforward calculation of the dynamical conductivity gives
\begin{eqnarray}
&\left.\begin{array}{c} \sigma_{xx}(\omega)\nonumber\\ i\sigma_{xy}(\omega)
\end{array}\right\}=i{\displaystyle\sigma_0
\frac{4\omega_c^2}{\pi^2}}
{\displaystyle\sum_{n,s,s'}\int\limits_0\limits^{\pi/2}\frac{dz}{\Delta_{ss'n}}
[f(\varepsilon _{s'n+1})-f(\varepsilon_{s,n})}]\\
&\times
{\displaystyle
\left[(\omega+i\Gamma
+\Delta_{ss'n})^{-1}\pm
(\omega+i\Gamma-\Delta_{ss'n})^{-1} \right]}\label{cond1}\\
&\times
\left|C^{2}_{sn}C^{1*}_{s'n+1}+ C^{3}_{sn}C^{4*}_{s'n+1}\right|^2 \,,
\nonumber
\end{eqnarray}
where  $\Delta_{ss'n}=\varepsilon_{sn}-\varepsilon_{s', n+1}$ is the level spacing,  $\omega_c=v\sqrt{2|e|\hbar B/c}$ is the cyclotron frequency, and $f(\varepsilon)=[\exp(\frac{\varepsilon-\mu}{T})+1]^{-1}$ is the
Fermi-Dirac function.  The integration over the Brillouin half-zone, $0<z<\pi/2$,  and the summation over the Landau number $n$ as well as  the bands $s, s'$ should be done in Eq. (\ref{cond1}). 

\begin{figure}[]
\resizebox{.52\textwidth}{!}{\includegraphics{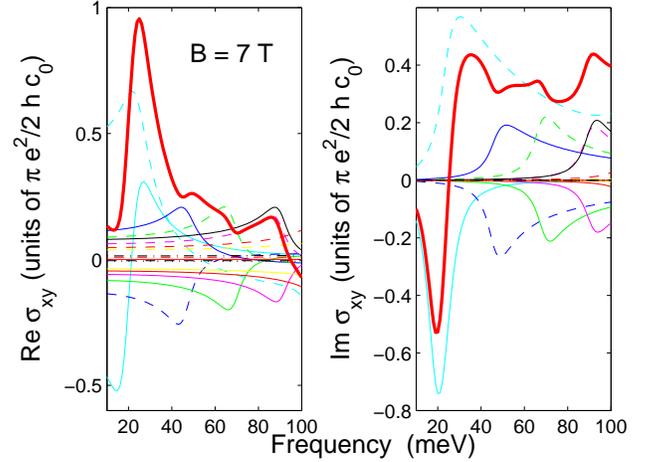}}
\caption{(Color online) Real and imaginary parts of the dynamical Hall conductivity at $B=$7 T (thick line) with the partial contributions of various electron transition are shown in the thin lines.}
\label{xy7}
\end{figure}

The selection rule $\Delta n=\pm1$ appears as a result of integration over $x$ and $x'$ in Eq. (\ref{cor}).
If the trigonal warping is taken into account, the selection rule is changed \cite{No,ST}. Let us notice
that the choice of the selection rule can be done
examining  the intensities of the cyclotron resonance lines. Our choice corresponds with observations in Refs.
\cite{OFS,CBN}.
\begin{figure}[]
\resizebox{.52\textwidth}{!}{\includegraphics{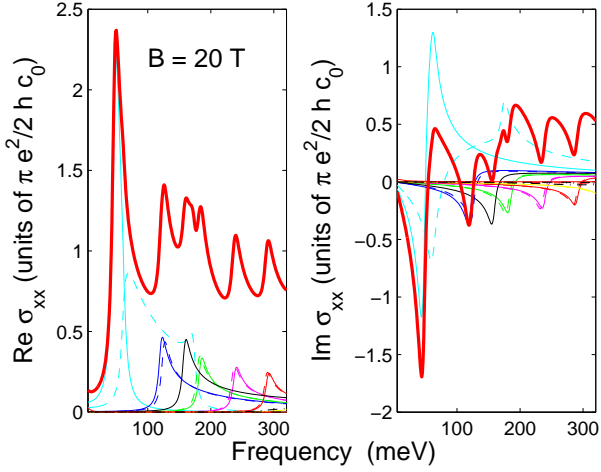}}
\caption{(Color online) Real and  imaginary parts of the longitudinal dynamical conductivity at $B=$20 T (thick line); the partial contributions of various electron transition are shown in the thin lines. Temperature T=0.1 meV is less than the level broadening $\Gamma=5$ meV.}
\label{xx20}
\end{figure}
\begin{figure}[]
\resizebox{.52\textwidth}{!}{\includegraphics{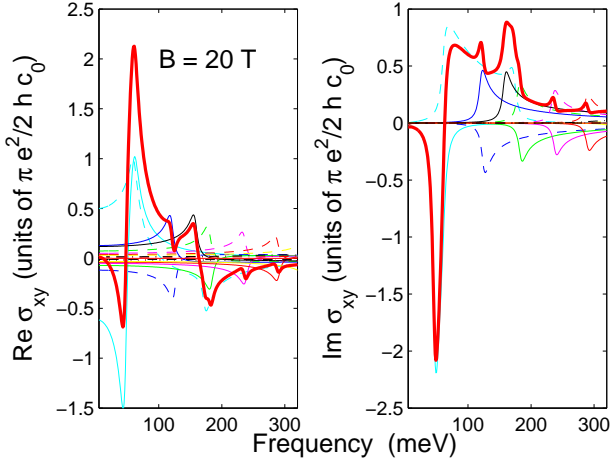}}
\caption{(Color online) Real and imaginary parts of the dynamical Hall conductivity at $B=$20 T (thick line) with the partial contributions of various electron transition shown in the thin lines.}
\label{xy20}
\end{figure}
The results for the longitudinal conductivity $\sigma_{xx}(\omega)$ are shown in  Figs. \ref{xx20} and \ref{xx7} for two values of the magnetic field in order to compare the effect of the field.
The dynamical Hall conductivity $\sigma_{xy}(\omega)$ shown in Figs. \ref{xy20} and \ref{xy7} describes the Faraday rotation \cite{CLW}. The formula  (\ref{cond1}) is valid in the collisionless limit,  when the relaxation rate is much less than the frequency, $\Gamma\ll\omega$.

In the calculations, we use the values  $\gamma_1=400$ meV, $\gamma_2=-10$ meV,$\gamma_5=5$ meV, 
$\Delta=45$ meV of Refs.\cite{PP,GAW}, and $\Gamma=5$ meV, $E_F=3 $ meV.

The conductivity units of 
$$\sigma_0=\frac{e^2}{4\hbar c_0}$$
have the simple meaning, being  the  graphene dynamic conductivity $e^2/4\hbar$ \cite{Kuz} multiplied by the number  $1/c_0$ of layers within the  distance unit in the $z-$direction.
One can see in Figs. \ref{xx20} and \ref{xx7}, that the value of conductivity calculated per one
graphite layer tends on average  to the  graphene universal conductance.
\begin{figure}[]
\resizebox{.52\textwidth}{!}{\includegraphics{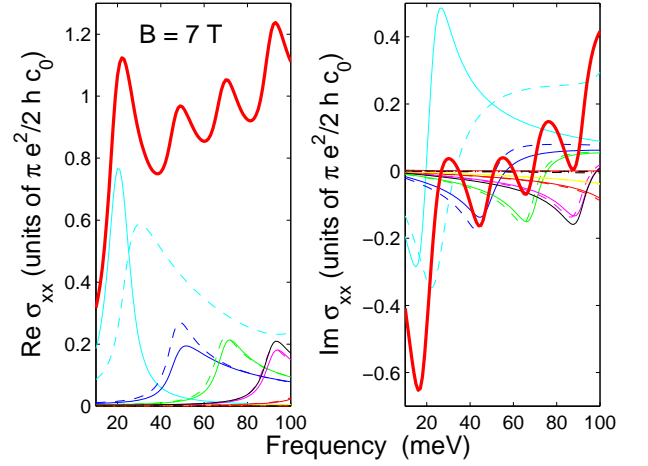}}
\caption{(Color online) The same as in Figs. \ref{xx20} but at $B=$7 T.}
\label{xx7}
\end{figure}

Let us analyze the spectroscopy of  the cyclotron resonances at $B=$ 20 T. The line at 292   meV is a doublet resulted from the electron transitions $|25\rangle\rightarrow|36\rangle\,$ and $|26\rangle\rightarrow|35\rangle $   near the $K$ point of the Brillouin zone (see Fig. \ref{disp20}). In the similar way,  the 240-meV line includes the $|24\rangle\rightarrow|35\rangle\,$ and $|25\rangle\rightarrow|34\rangle $ doublet splitted due to the electron-hole asymmetry. 
However, the  broad line at 165 meV involves  other transitions besides  the similar  $|23\rangle\rightarrow|34\rangle\,,$  $|24\rangle\rightarrow|33\rangle $ doublet at 184 meV. First is the transition
$|11\rangle\rightarrow|10\rangle $ (161 meV) near the $H$ point. 
Then, the transitions $|22\rangle\rightarrow|21\rangle $ produce 
the broad band. 
The high-frequency side  of the band (170 mev at $ H$) and the low-frequency side (70 meV, at the intersection of the  $|21\rangle- $level with the  Fermi level) contribute into  the  165-meV   and  50-meV lines, correspondingly. 
 The position of the broad band is very sensitive to the
variation of the tight-binding parameters and to the magnetic field.
The comparison of Fig. \ref{xx7} to Fig. \ref{xx20} shows that the
$|23\rangle\rightarrow|34\rangle\,$ line does not interfere with other
transitions at the weaker field. 
The main contribution into the sharp 50-meV line is resulted from
$|21\rangle\rightarrow|32\rangle $ transitions near the $K$-point.
The positions of the lines  for the fields in the range of 20 -- 30~ T agree very well  with observations of Refs. \cite{OFS,CBN}.
 We  do not correct the positions by a variation of the tight-binding parameters. Let us emphasize that
 the imaginary part of the dynamical conductivity is of the order of the real part.

The optical Hall conductivity $\sigma_{xy}(\omega)$ in the ac regime is shown in Figs. \ref{xy20} and   \ref{xy7}.  It is evident that the interpretation of the Faraday rotation governed by
the conductivity $\sigma_{xy}(\omega)$ (see  Fig. \ref{kerrang}) is much more complicated
in comparison with the longitudinal conductivity. The conductivities 
$\sigma_{xx}(\omega)$ and $\sigma_{xy}(\omega)$ allow us to calculate the reflectivity and the Faraday rotation as functions of frequency.

 Notice that the band structure shown in Fig. \ref{disp20} constrains to consider the (semi)metal-insulator transition while varying the carrier concentration and applying the magnetic field.  The phase transition induced by the electron-hole interaction has been discussed in  Refs. \cite{KTS,Kh,GGM}. As one can see in Fig. \ref{disp20}, the hole doping can decrease the Fermi energy. While the Fermi level appears between
 the $|22\rangle$ and  $|10\rangle$ levels, the insulator arises with a gap of 34 meV. 
  This phase transition does not involve the electron-electron interaction and it is resulted from the layered graphite structure   with the small electron dispersion of the zero mode in the $k_z$-direction. 

Author is thankful to A. Kuzmenko for discussions and corrections.
This work was supported by the Russian Foundation for Basic
Research (grant No. 10-02-00193-a) and by the SCOPES grant IZ73Z0$\_$128026 of the Swiss NSF. 
Author is grateful to the Max Planck Institute for the Physics of
Complex Systems for hospitality in Dresden.

\end{document}